# *Ab-initio* modeling of the short range order in Fe-N and Fe-C austenitic alloys

A.N. Timoshevskii , S.O. Yablonovskii

*Institute of magnetism of National academy of sciences and Ministry of education and science*

*of Ukraine, 36-B Vernadsky str., Kiev-142, Ukraine*

*Fax: 38 (044) 424 3075   E-mail:* tim@imag.kiev.ua

**Abstract**

In the present paper, we have studied atomic structure of nitrogenous austenite. High precision *ab-initio* calculation was utilized for the calculation of the pair potentials of interatomic interactions N-N in FCC Fe lattice. These potentials were used for the Monte Carlo modeling of the short range order in the Fe-N system. It was discovered that in FCC Fe lattice, nitrogen atoms might be partially ordered. In this case, atomic structure of nitrogenous austenite is characterized by availability of the $Fe_6N$ phase with the short range order over the N atoms located in the third coordination sphere.

**Introduction**

Most of the researchers relate the unique physical properties of the Fe-N austenite with the peculiarities of the short range order in this alloy. It is presumed that short range order in Fe-N and Fe-C austenitic alloys is essentially different, which, in its own turn, defines the difference in important physical properties like strength, plasticity and others. To describe short range order and obtain statistical and thermodynamic characteristics of the Fe-N and Fe-C austenites, it is necessary to obtain N-N and C-C interaction energies. In the pioneering papers, dedicated to such studies, the semi-phenomenological approaches based upon experimental data (for instance, temperature-concentration dependence of thermodynamic activity of the austenite interstitials [1]) were used. The undertaken investigations [2, 3 and 4] employed first order quasi-chemical approximation. First coordination sphere interactions obtained lead to a conclusion that C-C interaction has repulsion with the energy of $w_1$=0.083 eV [2] or $w_1$=0.04 eV [3] for Fe-C alloy. Similar estimations for N-N interaction energies were obtained for Fe-N alloy - $w_1$=0.04 eV [4] or $w_1$=0.02 eV [3]. It is possible to see that interaction energies absolute values differ significantly according to different authors. Obviously, this difference leads to different distribution of interstitials in the alloy. The possibility to study in more detail the parameters of the interatomic interaction is provided by Mössbauer spectroscopy [5-9]. Interatomic interaction energies C-C and N-N in the two coordination spheres were estimated with help of Monte-Carlo modeling [6-9]. Monte-Carlo modeling allows obtaining the distribution of the interstitial atoms detected by Mössbauer spectroscopy. The disadvantage of this approach is that, as a rule, it is possible to obtain the interaction only for two coordination spheres. Authors of the most papers, dedicated to the short range order study in nitrogenous austenite [5, 6, 8], drew the conclusion that nitrogen atoms are distributed according to $Fe_4N$ non-stoichiometric composition. Recently [10], the energies of formation for C-C and C-V (V-vacancy) pairs were obtained from the thermodynamic activity data analysis with the use of quasi-chemical approximation (QCA) of the statistical mechanics. The values of these energies were used in further Monte-Carlo modeling of the different Fe-C austenite concentrations. Obtained distributions of the atoms C and vacancies V were compared with known decompositions of Mössbauer spectra. Modeling results are in good agreement with $Fe_8C_{1-x}$ model proposed in [7]. Three types of iron atoms are distinguished in the frame of this model: $Fe_{10}$ – iron atom with one carbon atom as a closest neighbor in the first coordination sphere, $Fe_{0n}$ – iron atom with n carbon atoms only in the second coordination sphere and $Fe_{00}$ – iron atom, which does not contain carbon atoms in the two primary coordination spheres. It will be shown later, that our results are in good correspondence with $Fe_8C_{1-x}$ model. N-N atoms interaction was described in [11] with the help of Lennard-Jones pair potential with parameters obtained in the frame of quasi-harmonic approximation. The disadvantage of this approach is related to the supposition about an adequacy of the harmonic approximation and difficulties with obtaining of the information about electrochemical interaction potentials. Still, the values of the pair interaction energies were obtained for six coordination spheres ($w_1$=0.11, $w_2$=0.07, $w_3$=-0.06, $w_4$=0.01, $w_5$=0.03, $w_6$=0.02). As a result, the obtained decomposition up to 7$^{th}$ sphere indicates the formation of non-stoichiometric γ''-$Fe_8N_2$ phase, which is not observed experimentally. Therefore, until now, there is no clear common opinion about the distribution of nitrogen and carbon atoms in FCC iron lattice. Moreover, theoretical *ab-initio* short-range order modeling studies in these alloys are almost non-existent. So, the present paper is dedicated to the cluster decomposition [12] theoretical investigation of the Fe-N and Fe-C austenites.

**Results and discussion**

The essence of the approach used in the present investigation is the following. The energy of any atomic configuration in the crystal lattice of the binary alloy, where every site is occupied by the A or B atom, is possible to write in the form of the expansion $E=\sum V_\alpha P_\alpha$, where summation is done on all clusters. $V_\alpha$ – effective cluster interactions. $P_\alpha$ – product of the $s_i=(\pm 1)$ values, where index *i* runs about on the nodes of cluster. Cluster decomposition is the $P_\alpha$ basis expansion and it represents the generalization of the Ising model. $V_\alpha$ – structurally independent parameters, which are needed to be defined. The effectiveness of the cluster decomposition application is based upon the fact that, as a rule, it is possible to do with clusters of small dimension like, for instance, of 2, 3 or 4. In the present paper the pair approximation was used: clusters possess dimension of 2. Energy values for these clusters might be obtained from *ab-initio* calculations of the electronic structure for the set of the model ordered structures. Total energies of such structures together with known parameters give linear set of equations, which the unknown interaction values are found from. To solve this task, the high precision *ab-initio* FLAPW method (Wien2k package [13]) was used in the present paper. Exchange-correlation potential was calculated in the frame of gradient approximation (GGA) in accordance with Purdue-Burke-Ernzerhof model [14]. The number of k-points in the first Brillouin zone is 1000. The number of plane waves per atom in a basic set was equal to 160. This made it possible to ensure the precision for the total energy calculation of 0.001 eV. Calculations of the total energies of the ordered crystal structures were carried out taking into account full structural optimization, which includes lattice parameters and the unit cell atomic positions optimization.

Calculations have shown that in FCC iron lattice pair potentials for nitrogen-nitrogen (N-N) interatomic interaction depend essentially upon nitrogen concentration. In this work two sets of pair potentials of N-N interaction (one, which describes short range order in $Fe_4N$ nitride, and another one - for $FeN_{0.0625}$ austenite) have been calculated. One set of C-C pair potentials was also calculated for $FeC_{0.0625}$ austenite.

To define first set of potentials, the total energies of two ordered $Fe_8N_2$ structures, six $Fe_{32}N_4$ structures and one $Fe_{48}N_6$ structure have been calculated. According to Fe-N phase diagram [8] the formation of $Fe_4N$ nitride in the FCC iron matrix is possible in 12.5-25 at.% concentration range. Stoichiometry of the chosen model structures corresponds to this concentration interval. All calculations were carried out in spin-polarized ferromagnetic approach in order to describe ordering in the nitride, which is ferromagnetic itself. Calculation results for pair potentials and corresponding coefficients in the total energy equations for all structures $E=A_0E_0+A_1E_1+\sum c_i v_i$ are given in Table 1. To define the set of potentials of the $FeA_{0.0625}(A=N,C)$ austenites, total energies of the five ordered $Fe_{32}A_2(A=N,C)$ structures and two $Fe_{48}A_3(A=N,C)$ ones were calculated. As these austenites are paramagnetic, calculations were carried out in non-spin-polarized approximation. Calculated pair potentials and corresponding coefficients in the total energy equations for all structures $E=A_0E_0+\sum c_i v_i$ are given in Table 2. Two sets of pair potentials for nitride and nitrogenous austenite are essentially different. In the case of nitride, nitrogen atoms attraction is observed in the second coordination sphere, while in the case of austenite – in the third one, which must lead to fundamentally different short range order in these materials. Potentials for nitrogenous and carbon austenite are also different. The C-C interaction is characterized by the repulsion in the third coordination sphere.

Table 1. Coefficients of the set of equations $c_i$, total energies of the model structures $E$ and pair potentials of atomic interactions of the nitrogen atoms $v_i$ (eV) for $Fe_4N$ nitride.

| Type of structures | n | $A_0$ | $A_1$ | $c_1$ | $c_2$ | $c_3$ | $c_4$ | $E_n$ |
|---|---|---|---|---|---|---|---|---|
| $Fe_8N_2$ | 1 | 1 | 0 | 0 | 1/2 | 1 | 1/2 | -35007.361 |
|  | 2 | 1 | 0 | 1/2 | 1/2 | 0 | 1/2 | -35007.293 |
| $Fe_{48}N_6$ | 3 | 0 | 1 | 0 | 1/24 | 0 | 7/12 | -34820.995 |
| $Fe_{32}N_4$ | 4 | 0 | 1 | 0 | 1/4 | 0 | 1/4 | -34821.009 |
|  | 5 | 0 | 1 | 0 | 3/16 | 0 | 3/8 | -34820.992 |
|  | 6 | 0 | 1 | 0 | 0 | 0 | 3/4 | -34820.998 |
|  | 7 | 0 | 1 | 0 | 0 | 3/8 | 0 | -34820.995 |
|  | 8 | 0 | 1 | 3/16 | 0 | 0 | 0 | -34820.963 |
|  | 9 | 0 | 1 | 1/16 | 0 | 1/8 | 0 | -34820.986 |
|  |  | $E_0$ | $E_1$ | $v_1$ | $v_2$ | $v_3$ | $v_4$ |  |
|  |  | -35007.368 | -34820.998 | 0.169 | -0.022 | 0.014 | 0.004 |  |

Table 2. Coefficients of the set of equations $c_i$, total energies of the model structures $E$ and pair potentials of atomic interactions of the nitrogen atoms $v_i$ (eV) for $FeA_{0.0625}$(A=N, C) alloys.

| Type of structures | n | $A_0$ | $c_1$ | $c_2$ | $c_3$ | $c_4$ | $c_5$ | $E^N_n$ | $E^C_n$ |
|---|---|---|---|---|---|---|---|---|---|
| $Fe_{32}A_2$ | 1 | 1 | 1/32 | 0 | 0 | 0 | 2/32 | -34727.814 | -34699.455 |
|  | 2 | 1 | 0 | 0 | 0 | 0 | 0 | -34727.817 | -34699.460 |
|  | 3 | 1 | 0 | 0 | 1/16 | 0 | 0 | -34727.818 | -34699.458 |
|  | 4 | 1 | 0 | 0 | 0 | 1/8 | 0 | -34727.813 | -34699.453 |
| $Fe_{48}A_3$ | 5 | 1 | 0 | 1/48 | 0 | 0 | 0 | -34727.810 | -34699.453 |
|  | 6 | 1 | 0 | 0 | 2/48 | 2/48 | 2/48 | -34727.817 | -34699.457 |
|  |  | $E_0$ | $v_1$ | $v_2$ | $v_3$ | $v_4$ | $v_5$ |  |  |
| N-N | 1 | -34727.817 | 0.131 | 0.298 | -0.013 | 0.030 | -0.024 |  |  |
| C-C | 2 | -34699.460 | 0.177 | 0.338 | 0.028 | 0.056 | -0.007 |  |  |

Different thermodynamic characteristics of the nitride and austenitic $FeN_{0.0625}$ alloy were calculated in the broad temperature interval by the Monte-Carlo method using two sets of the obtained potentials. The changes in nitrogen atomic positions were carried out with hopping probabilities according to Metropolis. The cell with $24^3$ atomic positions was used. Temperature step was 100 K with fixed nitrogen concentration. Short range order calculation was conducted with the decrease of temperature. Equilibrium configuration of the previous calculation was used as a starting one. Phase transformation points were found from the maximum location on the heat capacity versus temperature curve. Short range order modeling in $FeN_x$ alloy with primary set of potentials was conducted for concentrations x=0.05, 0.10, 0.15, 0.25. Fig. 1 shows concentration dependencies of the number of iron atoms of different type in alloy at temperature T=800 K. As it can be seen at x=0.25 nitrogen concentration ($Fe_4N$ nitride), only atoms $Fe_2$-180º ($Fe_2$ atom possesses nitrogen atoms, which are under the angle 180º) and atoms $Fe_0$ exist.

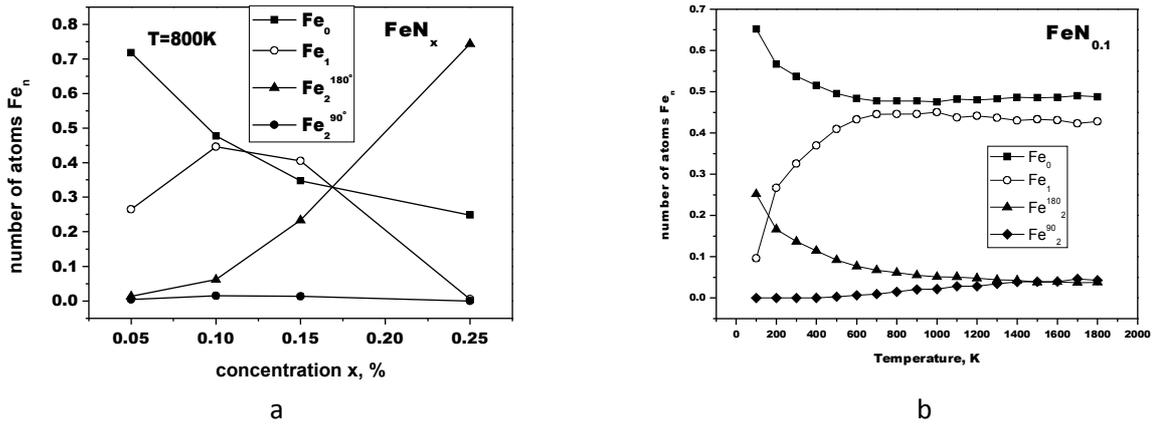

Fig. 1. a) Concentrational dependence of the number of different iron atoms in $FeN_x$ (x=0.05, 0.1, 0.15) alloy and $Fe_4N$ nitride; b)Temperature dependence of the number of different iron atoms in $FeN_{0.1}$ alloy.

It is an evidence of the $Fe_4N$ nitride phase formation. Calculations have shown that this phase forms at 800 K, which correlates with the experimental value of the nitride formation temperature (923K). Further, the Monte-Carlo modeling of the short range order in austenitic alloys with smaller nitrogen concentrations was carried out with the help of the same interaction pair potentials. Calculations have also shown that at 800 K iron atoms of different kind exist in these alloys. For instance, there are atom $Fe_1$, which possesses one nitrogen atom, and atom $Fe_2$, which possesses two nitrogen atoms under the angle 90°. However, at modeling for alloys of all x=0.05,0.10,0.15 concentrations, starting from certain temperature, that decreases with the decrease of nitrogen concentration, the ordering takes place, which is characteristic for nitride formation. It contradicts the experimental data, as it is well-known that in the austenitic alloys no nitride formation is observed at such nitrogen concentrations. Thus, the set of potentials given in Table 1 does not describe the short range order in austenite alloy correctly.

Monte-Carlo modeling of the short range order in nitrogenous and carbon austenites employed pair potentials obtained with the help of model structures $Fe_{32}A_2$ and $Fe_{48}A_3$ (A=N, C), which are actually modeling $FeA_{0.0625}$(A=N,C) alloys. Calculations have shown that in the case of nitrogenous austenite no $Fe_4N$ nitride formation is observed at the temperature decrease. Fig. 2 shows the results of short range order modeling in $FeA_{0.0625}$(A=N,C) alloys. The temperature dependence of the number of iron atoms of different type ($Fe_1^4$, $Fe_1^{5-12}$ and $Fe_0$) was calculated. $Fe_1^4$ atoms possess one impurity atom and 4 iron atoms in the first coordination sphere. Correspondingly, $Fe_1^{5-12}$ atoms possess from 5 to 12 iron atoms of $Fe_1$ type. For the $FeN_{0.0625}$ nitrogenous austenite the increase in number of $Fe_1^{5-12}$ atoms with the decrease of temperature (Fig. 2a) is an evidence of the partial ordering of the nitrogen atoms in the FCC iron lattice. For the $FeC_{0.0625}$ carbon austenite (Fig. 2b), starting from 400 K the sharp decrease of the amount of $Fe_1^{5-12}$ atoms is observed, which is an evidence of carbon atoms disordering in FCC iron lattice. Analysis of the results has shown that nitrogen atoms are situated mainly in the third coordination sphere in respect to each other, while carbon atoms – in the 5$^{th}$ and 6$^{th}$ coordination spheres. It means that, in the case of nitrogenous austenite with the temperature decrease, the $Fe_6N$ octahedra, consisting of $Fe_1$ atoms, unite but do not possess common iron atoms. In the case of the carbon austenite the carbon atoms are distributed at the maximal distance from each other. Actually, the phase of the $Fe_6N$ stoichiometry with short range order of the nitrogen atoms in the third coordination sphere is formed in the FCC iron lattice of the nitrogenous austenite. The obtained results lead to the conclusion that in this case it is impossible to form long range order over the nitrogen atoms. $Fe_1^4$ atoms are situated on the surface of this phase. As it can be seen in Fig. 2a, the number of these atoms decreases sharply starting at the temperature of formation of the short range order over the nitrogen and formation of the phase with $Fe_6N$ stoichiometry. For the carbon austenite with the temperature decrease, the number of $Fe_1^4$ iron atoms increases (Fig. 2b), which might help to explain smaller solubility of carbon atoms (9 at. %) comparing to nitrogen (10.5 at. %) in FCC iron matrix. It has to be noted that the number of $Fe_0$ atoms does not depend upon temperature.

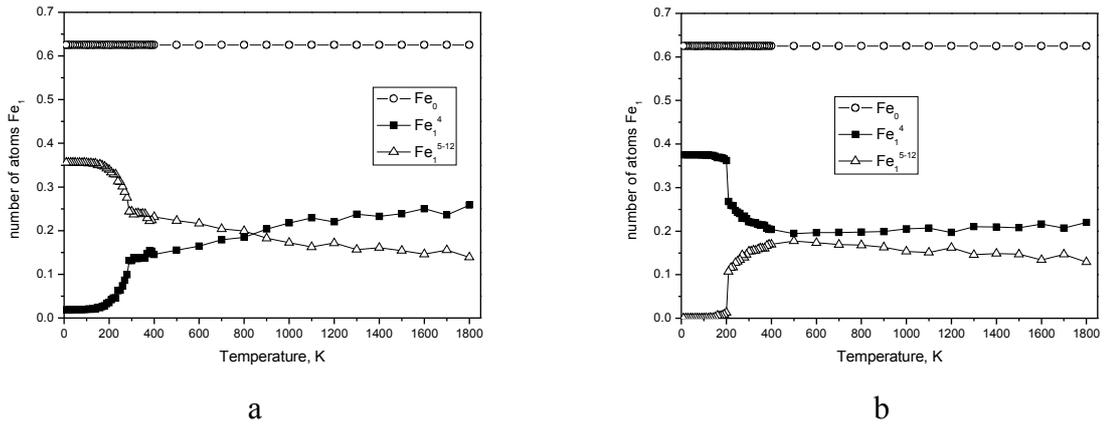

Fig. 2. Temperature dependence of the number of different iron atoms in a) $FeN_{0.0625}$, b) $FeC_{0.0625}$ alloys.

The results of the Monte-Carlo modeling allow to single out the fragment of the FCC lattice, which demonstrates the distribution of the nitrogen atoms in the third coordination sphere. Fig. 3a represents the scheme of the $Fe_{108}N_9$ structure, which might be a good model for the nitrogenous austenite. $Fe_1$ atoms are marked with the dark color around each nitrogen atom. These atoms form the $Fe_6N$ octahedra in the FCC lattice. Figure 3b shows $Fe_{32}C_2$ structure, which appeared to be energy favorable amongst all calculated structures. In this structure carbon atoms are situated in $6^{th}$ coordination sphere. It corresponds to $Fe_8C_{1-x}$ model proposed in [7] and might be a good model for the calculation of the hyperfine interactions.

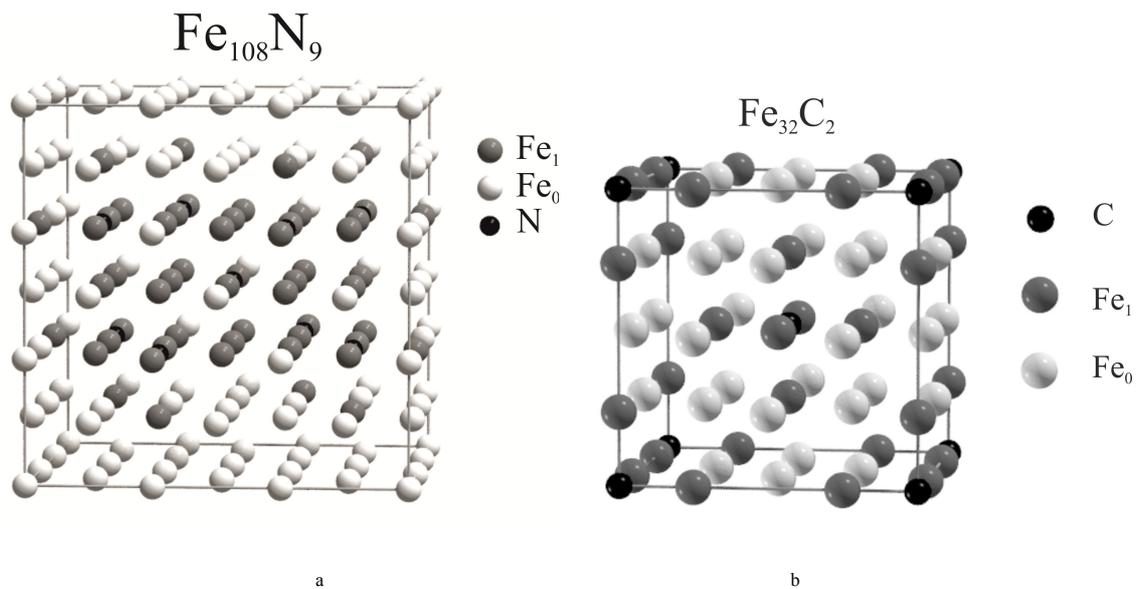

Fig. 3. a) Structure $Fe_{108}N_9$, simulating the nitrogen atoms distribution in the third coordination sphere of FCC iron lattice. b) Structure $Fe_{32}C_2$, simulating the nitrogen atoms distribution in the sixth coordination sphere of FCC iron lattice.

**Conclusion**

It was found that the nitrogen atoms in the nitrogenous austenite can be partially ordered. In this case, its atomic structure is characterized by the presence of the $Fe_6N$ phase with the short range order over the atoms of nitrogen, which are located in the third coordination sphere.

Calculations have shown that, contrary to nitrogen, carbon atoms in iron FCC lattice repulse and settle at maximum possible distances from each other.